\documentstyle[epsfig,12pt,preprint,tighten,aps]{revtex}
\begin{document}

\draft

\title{\rightline{{\tt May 2000}}
\rightline{{\tt UM-P-00/021}}
\rightline{{\tt RCHEP-00/003}}
\ \\
Discovering mirror particles at the Large Hadron Collider and the implied
cold universe}
\author{A. Yu.\ Ignatiev and R. R. Volkas}
\address{School of Physics\\
Research Centre for High Energy Physics\\
The University of Melbourne\\
Victoria 3010 Australia\\
(sasha@physics.unimelb.edu.au, r.volkas@physics.unimelb.edu.au)}

\maketitle

\begin{abstract}

The Mirror Matter or Exact Parity Model sees every standard particle,
including the physical neutral Higgs boson, paired with a parity partner. The
unbroken parity symmetry forces the mass eigenstate Higgs bosons to be maximal
mixtures of the ordinary and mirror Higgs bosons. Each of these mass
eigenstates will therefore decay $50\%$ of the time into invisible mirror
particles, providing a clear and interesting signature for the Large Hadron
Collider (LHC) which could thus establish the existence of the mirror world.
However, for this effect to be observable the mass difference between the two
eigenstates must be sufficiently large. In this paper, we study cosmological
constraints from Big Bang Nucleosynthesis on the mass difference parameter. We
find that the temperature of the radiation dominated (RD) phase of the
universe should never have exceeded a few 10's of GeV if the mass difference
is to be observable at the LHC. Chaotic inflation with very inefficient
reheating provides an example of how such a cosmology could arise. We conclude
that the LHC could thus discover the mirror world and simultaneously establish
an upper bound on the temperature of the RD phase of the universe.

\end{abstract}

\section{Introduction}

The Mirror Matter or Exact Parity Model (EPM) sees every standard particle paired
with a parity partner. This idea was first mentioned by Lee and Yang in their
seminal paper on parity violation \cite{ly} as a way to retain the full Poincar\'e
Group as a symmetry of nature despite the $V-A$ character of weak interactions.
Some follow up work was performed in the ensuing decades on some aspects of the
mirror matter hypothesis \cite{kob}. In 1991, the idea was independently
rediscovered and the full gauge theory constructed for the first time \cite{flv1}.
Shortly thereafter the EPM was extended to include nonzero neutrino masses and
mixings and applied to the solar, atmospheric and LSND anomalies \cite{flv2,flv3}.
The EPM can also alter standard Big Bang cosmology in interesting ways, through the
possible identification of some dark matter with mirror matter, and through
modifications of Big Bang Nucleosynthesis (BBN) \cite{fv}.

The ordinary and mirror particle sectors can interact in a number of ways. The first
is through gravitation, with immediate consequences for the dark matter problem and
astrophysics. Non-gravitational interactions can be induced through the mixing of
colourless and neutral particles with their mirror counterparts. Neutrinos, the
photon, the $Z$ boson and the physical neutral Higgs boson can mix with the
corresponding mirror states. Coloured and/or electrically charged particles are
prevented from mixing with their mirror analogues by colour and electric charge
conservation laws.

The purpose of this paper is to study the Higgs boson sector of the EPM. It has
been previously noted that the mass eigenstate physical Higgs bosons must be
maximal mixtures of the underlying ordinary and mirror states because of the
unbroken parity symmetry \cite{flv1,flv2}. Each mass eigenstate will therefore
decay $50\%$ of the time into invisible mirror particles, providing a striking
experimental signature in principle.
The production cross-section for such Higgs bosons would be 1/2 of that
of the standard Higgs boson of the same mass. This is a very simple and
important
observation, because it provides a clear way to experimentally establish the
existence of the mirror world \footnote{From the recent results of the 
L3 Collaboration \cite{L3}
we can establish a lower bound of about 65 GeV for a Higgs boson with these
properties.}.
 In recent years there has been a strong focus on
using the neutrino anomalies as a way to discover mirror 
matter \cite{flv2,flv3,fv}. Neutrino oscillation physics 
certainly does provide a very interesting way to
garner experimental evidence for mirror matter, or to at least constrain the model
(if one is being pessimistic). However, the terrestrial neutrino phenomenology of
the EPM is similar to that of pseudo-Dirac neutrinos \cite{pd}, so complementary
information would be useful. The Higgs boson sector is one potentially important
way to obtain this information. (The mixing of ortho-positronium with
mirror-ortho-positronium is another \cite{fg}.)

The strength of Higgs boson mixing with its mirror partner is controlled by an {\it
a priori} independent dimensionless parameter $\lambda_{HH'}$. The mass splitting
between the mass eigenstate Higgs bosons is proportional to this same parameter.
Standard cosmology, through BBN, can be used to constrain $\lambda_{HH'}$ and hence
the Higgs boson mass splitting also \cite{cf}. In this paper, we will demonstrate
that the temperature of the radiation dominated (RD) phase of the
universe
should never have exceeded a few tens of GeV if the mass splitting is to be substantial (of
order 1 GeV). Chaotic inflation with very inefficient reheating is {\it an
example} of how such a cold cosmology could arise. Remarkably, the Large Hadron
Collider (LHC) could thus discover the mirror world as a byproduct of its Higgs
boson search programme, and simultaneously establish an upper bound on the
temperature of the RD phase of the universe.

\section{The Higgs boson sector and cosmological constraints}

Consider a minimal Higgs boson sector for the EPM. It contains the standard
Higgs doublet $\phi$, transforming as a ${\bf 2}(1)$ representation under the
electroweak gauge group SU(2)$ \otimes$U(1)$_Y$. It also contains a mirror
Higgs doublet $\phi'$ which transforms as a ${\bf 2}(1)$ representation under
the mirror electroweak gauge group SU(2)$' \otimes$U(1)$'_Y$. The standard
doublet $\phi$ is a singlet under the mirror gauge group, while $\phi'$ is a
singlet under the ordinary gauge group. Under the discrete parity symemtry,
$\phi \leftrightarrow \phi'$.

We focus on the Higgs potential in this paper.
It is very simply given by
\begin{equation}
V = \lambda_+ ( \phi^{\dagger} \phi + \phi'^{\dagger} \phi' - 2 v^2 )^2
+ \lambda_- ( \phi^{\dagger} \phi - \phi'^{\dagger} \phi' )^2.
\label{V}
\end{equation}
In the $\lambda_{\pm} > 0$ region of parameter space, the vacuum is clearly given by
\begin{equation}
\langle \phi \rangle = \langle \phi' \rangle = 
\left( \begin{array}{c} 0 \\ v \end{array} \right).
\end{equation}
In this region of parameter space, the parity or mirror symmetry is respected by the
vacuum because of
the equality between the vacuum expectation values (VEVs) of the ordinary and mirror
Higgs doublets.\footnote{A parity {\it breaking} global minimum of this Higgs
potential
can be found in another region of parameter space \cite{fl}.} Going to unitary
gauge and
shifting the neutral components as
per $\phi^0 = v + H/\sqrt{2}$ and $\phi'^0 = v + H'/\sqrt{2}$ we see from Eq.(\ref{V})
that the mass eigenstates are
\begin{equation}
H_{\pm} = \frac{ H \pm H' }{\sqrt{2}}
\label{Hpm}
\end{equation}
with masses given by
\begin{equation}
m^2_+ = 8 \lambda_+ v^2\quad \text{and}\quad m^2_- = 8 \lambda_- v^2
\end{equation}
respectively. We therefore see that the mass splitting
\begin{equation}
\Delta m \equiv m_+ - m_- = (\lambda_+ - \lambda_-) \frac{8 v^2}{m_+ + m_-}
\end{equation}
is controlled by the parameter
\begin{equation}
\lambda_{HH'} \equiv \lambda_+ - \lambda_-.
\end{equation}
 From Eq.(\ref{V}) we also see that the coefficient of the $HH'$ mixing term,
$4 \lambda_{HH'} v^2$, is proportional to the same parameter. In addition, 
the coefficient of the quartic
coupling term $\phi^{\dagger} \phi \phi'^{\dagger} \phi'$ is $2\lambda_{HH'}$.

It is clear that each mass eigenstate physical neutral Higgs boson $H_{\pm}$ decays
$50\%$ of the time into ordinary particles and $50\%$ of the time into mirror, and
hence invisible, particles. The total decay rate of $H_+$ or $H_-$ is the same as that
for a SM physical neutral Higgs boson of the same mass. Note also that each mass
eigenstate couples to ordinary particles with strength reduced by $1/\sqrt{2}$
compared to the coupling of the standard Higgs boson to those same particles
\cite{flv1,flv2}.

We now turn to cosmological constraints from BBN on $\lambda_{HH'}$, or
equivalently, $\Delta m$. BBN does not allow the mirror plasma to be in thermal
equilibrium with the ordinary plasma during the relevant epoch. The parameters
controlling the mixing of colourless and neutral particles with their mirror
partners must therefore obey upper bounds, assuming that the standard theory of BBN
is correct. The derivations of these bounds for the photon and neutrino systems
have been described elsewhere \cite{fv,photon}.

The Higgs system situation was briefly discussed in Ref.\cite{cf} and will
be fully explored
here. There are two different epochs to consider: (i) temperatures $T
\stackrel{>}{\sim} 100$'s of GeV, where the Higgs bosons exist as real particles
in the plasma, and (ii) the opposite limit where they do not.

Epoch (i) was considered
in Ref.\cite{cf}. The physics is very simple. Suppose that no mirror
particles exist in the
plasma to begin with. We then have to ensure that processes driven by $\lambda_{HH'}$
do not bring the mirror Higgs bosons, and hence all other mirror particles, 
into thermal equilibrium. During epoch (i), the electroweak symmetry is presumably
restored, so the relevant term is $2 \lambda_{HH'} \phi^{\dagger} \phi \phi'^{\dagger}
\phi'$ from the Higgs potential. By dimensional analysis, the rate for $\phi \phi \to
\phi' \phi'$ scattering will be approximately given by
$(\lambda_{HH'})^2 T$. Requiring that this be less than the expansion rate of the
universe
$\simeq \sqrt{g} T^2/m_P$, where $g \simeq 100$ is the effective number of
massless degrees of freedom and $m_P$ is the Planck mass, we find the bound
\begin{equation}
\lambda_{HH'} \stackrel{<}{\sim} 10^{-8} \sqrt{\frac{m_\phi}{100\text{GeV}}}.
\label{highTbound}
\end{equation}
This bound is obtained by setting the temperature $T$ to be about $m_\phi$ in order to
get the most restrictive condition, where $m_\phi \simeq 100$'s of GeV is the Higgs
mass in the symmetric phase. This is clearly a severe bound. If the assumptions behind
its derivation were unassailable, then the EPM would have a significant fine-tuning
problem: why is $\lambda_{HH'}$ so small? It has been observed that the
supersymmetric extension of the EPM yields $\lambda_{HH'} = 0$
\cite{d}. While this is of
interest, we will look for an alternative solution, because $\lambda_{HH'} = 0$
eliminates
the chance for the LHC to discover the mirror world. For a sufficiently small
$\Delta m$, the
ordinary particles from which the Higgs bosons are produced yield a coherent
superposition of $H_+$ and $H_-$ which is precisely the ordinary Higgs boson $H$.
The Higgs physics of the EPM is then indistinguishable from that of the Standard
Model. According to Ref.\cite{gninenko}, the LHC is expected to measure the Higgs
boson mass with an accuracy of roughly $1\%$. This means that $\lambda_{HH'}$ must
be larger than about $0.01$ for the mass difference between $H_+$ and $H_-$ to be
observable.

So, let us suppose that the radiation dominated phase of the universe was
never hot enough for Higgs bosons to exist as
real particles in the plasma! The bound of Eq.(\ref{highTbound}) is then irrelevant.
Such a ``cold universe'' can be produced, for example, by inefficient reheating after
inflation. We will discuss this further in the next section.

For $T \ll m_\phi$, mirror particles can be brought into thermal equilibrium via the
$f \overline{f} \to f' \overline{f'}$ process mediated by
virtual Higgs boson exchange as depicted in the Figure. We will also take $T$ to be
less than
the temperature for the electroweak phase transition (about 100 GeV), so we are in the
broken phase. The rate is given roughly by
\begin{equation}
\Gamma \sim h_f^4 \lambda^2_{HH'} \frac{v^4 T^5}{m^8_\phi},
\label{rate}
\end{equation}
where $h_f = m_f/v$ is the Yukawa coupling constant for the fermions and mirror
fermions in the initial and final states. For $T \ll 100$ GeV, top quarks are not a
significant component of the plasma, so the bottom quark $b\overline{b} \to
b'\overline{b'}$ process will
dominate all others. We therefore set $h_f = h_b = m_b/v$. The condition
that $\Gamma$ is always less than the expansion rate then implies that
\begin{equation}
\lambda_{HH'} \stackrel{<}{\sim} \frac{m^4_{\phi}g^{1/4}}{m_b^2 \sqrt{m_P T^3}}
\simeq 0.1\left( \frac{T_{\text{max}}}{\text{GeV}}
\right)^{-\frac{3}{2}}
\left( \frac{m_{\phi}}{200\text{GeV}}\right)^4.
\label{lowTbound}
\end{equation}
which is most restrictive for the highest temperature $T_{\text{max}}$ we hypothesise
the radiation dominated phase of the universe to
reach. 
If we require
that the Higgs boson mass difference be observable at the LHC ($\lambda_{HH'}
\stackrel{>}{\sim} 0.01$) then $T_{\text{max}}$ cannot be higher than
a few tens of GeV. Furthermore, if $T_{\text{max}}$ does not exceed a few GeV
then $\lambda_{HH'}\sim 1$ is allowed \footnote{If $T_{\text{max}}$ is below $m_b \simeq 4.4$ GeV,
then charmed quarks
and tau leptons should be used instead of bottom quarks.}.

The cold universe hypothesis simultaneously remedies the $\lambda_{HH'}$ fine-tuning
problem, and allows a large mass splitting between $H_+$ and $H_-$. Remarkably, the
LHC could simultaneously discover the mirror world and produce strong evidence that
the RD phase of the universe was never hotter than a few 
tens of GeV!

\section{Cold cosmology?}

We will now make a few remarks about how such a cold cosmology could be constructed.
As well as providing a low $T_{\text{max}}$, the cosmological model would also have to
explain why the early universe was predominantly composed of ordinary matter in the
first place. A universe with an ordinary and mirror plasma in thermal equilibrium with
each other is ruled out by BBN. Phrased another way, we must require that the
temperature $T'$ of the mirror plasma be less than about half of the temperature of
the ordinary plasma during the BBN epoch in order for the expansion rate of the
universe to not be too high.

The $T'$ issue has already been addressed in the literature through inflationary
models \cite{inf,bv}. As an example, Ref.\cite{bv} introduces an inflaton $\sigma$
and a mirror inflaton
$\sigma'$ with the potential
\begin{equation}
U = \frac{1}{2} m^2_{\sigma} (\sigma^2 + \sigma'^2)
\end{equation}
in the context of the chaotic inflationary paradigm (see, e.g. \cite{linde}).
They then suppose that the
chaotic initial conditions set up $\sigma' \ll \sigma$ by chance.\footnote{This
important idea
illustrates how a cosmology which is asymmetric between the ordinary and mirror
sectors can arise despite the identical microphysics: exploit fluctuations.}
The equations of
motion derived from $U$ then show that $\sigma'/\sigma$ remains constant during the
inflationary phase. This means that the $\sigma'$ field will begin oscillating about
$\sigma' = 0$ while $\sigma$ is still driving inflation. Assuming further that
$\sigma'$ $(\sigma)$ couples only to mirror (ordinary) particles, the reheated mirror
plasma created by the decays of $\sigma'$ gets diluted by the inflationary expansion
that is still occurring. When $\sigma$ subsequently ceases driving inflation, it then
produces a reheated ordinary plasma that has a much higher temperature than the
diluted mirror plasma. All we need to further postulate is very inefficient reheating
to produce the required cold universe. One will ultimately also need a baryogenesis
mechanism that can work at such low temperatures. Some proposals already exist in the
literature, including the exploitation of the pre-heating process for this
purpose \cite{lowT}.

The above is but an example of how a cold cosmology with asymmetric temperatures for
the ordinary and mirror plasmas might arise. Many other issues need to be addressed,
including the origin of the inflaton potential, the precise mechanism of reheating,
and whether a substantial (but still subdominant) amount of mirror matter can be
produced in addition to the ordinary matter during re- or pre-heating (as would be
needed for mirror dark matter purposes). For the moment, our focus should be on the
interesting and simple Higgs physics of the EPM. If the LHC discovers a large mass
splitting between $H_+$ and $H_-$, then this terrestrially obtained data will provide
good motivation for further work in cosmological model building.

\section{Conclusions}

The Exact Parity or Mirror Model predicts some simple and interesting Higgs
physics. There will be two physical neutral Higgs boson mass eigenstates, each
with a $50\%$ invisible width. This would be a remarkable way to discover
mirror particles. A detectable mass splitting between the two eigenstates
would strongly suggest that the radiation dominated phase of the
universe was never hotter than, say, a few tens of GeV. This would in turn be
interesting information for cosmological model builders.

\acknowledgments{We would like to thank Nicole Bell and Sergei Gninenko for
discussions and Robert
Foot for comments on a draft manuscript. A.I. is grateful to D.Grigoriev and
M.Shaposhnikov for helpful discussions. This work was supported by the
Australian Research Council.

\pagestyle{empty}   
\newpage
\centerline{\Large FIGURE}

\vspace{3cm}

\epsfig{file=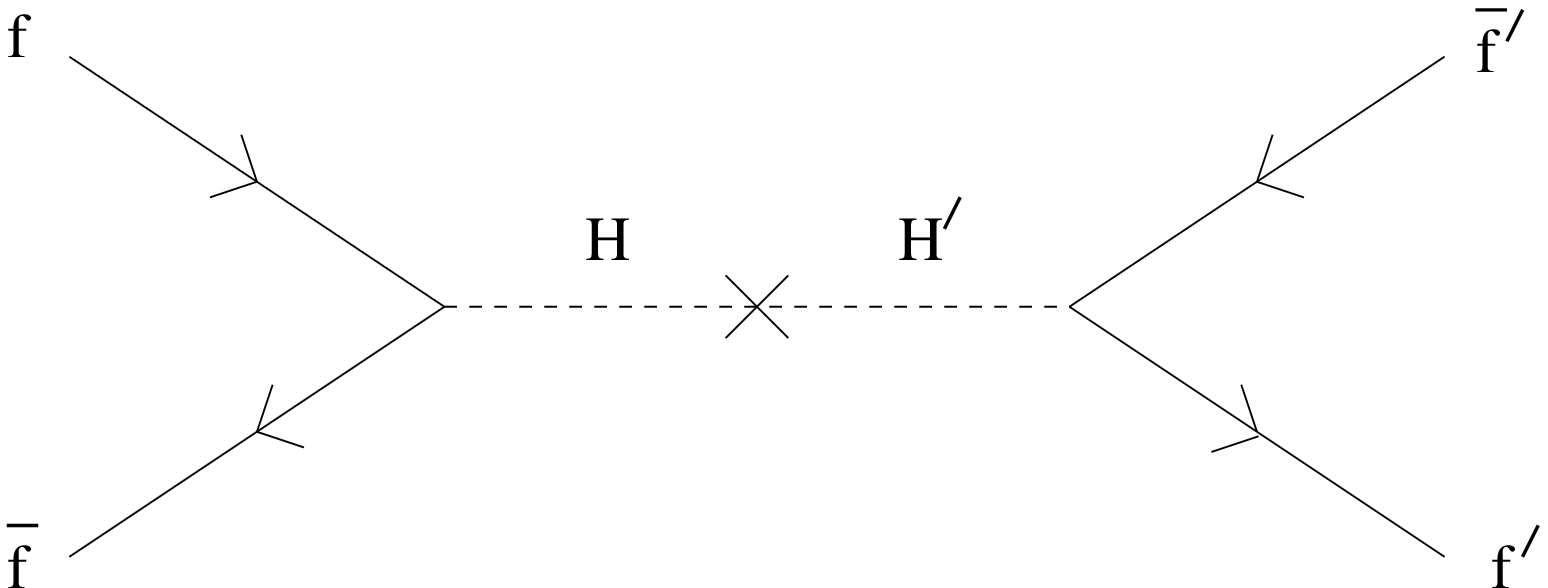,width=15cm}

\vspace{1cm}
Diagram of the process $f \overline{f} \to f' \overline{f'}$ mediated by
Higgs--mirror-Higgs boson mixing.

\end{document}